\documentclass[sigconf, nonacm]{acmart}

\newcommand{\dcaption}[2]{\caption[#1]{#1. \textmd{\small#2}}}

\usepackage{enumerate}
\usepackage{subfig}
\usepackage{hyperref}

\usepackage{graphicx}
\usepackage{fixmath}
\usepackage{tikz}
\usetikzlibrary{tikzmark,calc}

\begin{document}
\title{An efficient method to automate tooth identification and 3D bounding box extraction from Cone Beam CT Images}

\author{Ignacio Garrido Botella}
\affiliation{%
  \institution{Dezzai}
  \city{Madrid}
  \state{Spain}
}
\email{ignacio.garrido@dezzai.com}

\author{Ignacio Arranz Águeda}
\affiliation{%
  \institution{Dezzai}
  \city{Madrid}
  \state{Spain}
}
\email{ignacio.arranz@dezzai.com}

\author{Juan Carlos Armenteros Carmona}
\affiliation{%
 \institution{Dezzai}
  \city{Madrid}
  \state{Spain}
}
\email{juancarlos.armenteros@dezzai.com}

\author{Oleg Vorontsov}
\affiliation{%
  \institution{Dezzai}
  \city{Madrid}
  \state{Spain}
}
\email{o.vorontsov@dezzai.com}

\author{Fernando Bayón Robledo}
\affiliation{%
  \institution{Dezzai}
  \city{Madrid}
  \state{Spain}
}
\email{f.bayon@dezzai.com}

\author{Evgeny Solovykh}
\affiliation{%
\institution{LLC FDLAB}
\city{Moscow}
\state{Russian Federation}
}
\email{solovykh75@gmail.com}

\author{Obrubov Aleksandr Andreevich}
\affiliation{%
\institution{Radiology department of the Central Research Institute of Dentistry and Maxillofacial Surgery}
\city{Moscow}
\state{Russian Federation}
}
\email{obrubov.a@outlook.com}

\author{Adrián Alonso Barriuso}
\affiliation{%
  \institution{Dezzai}
  \city{Madrid}
  \state{Spain}
}
\email{a.alonso@dezzai.com}



\begin{abstract}

Accurate identification, localization, and segregation of teeth from Cone Beam Computed Tomography (CBCT) images are essential for analyzing dental pathologies. Modeling an individual tooth can be challenging and intricate to accomplish, especially when fillings and other restorations introduce artifacts. This paper proposes a method for automatically detecting, identifying, and extracting teeth from CBCT images. Our approach involves dividing the three-dimensional images into axial slices for image detection. Teeth are pinpointed and labeled using a single-stage object detector. Subsequently, bounding boxes are delineated and identified to create three-dimensional representations of each tooth. The proposed solution has been successfully integrated into the dental analysis tool Dentomo.

\end{abstract}

\maketitle

\section{Introduction}
\label{sec:intro}

In the last few years, there has been a significant push towards automating disease diagnosis in clinical settings \cite{ai_for_med}. Advancements in medical imaging techniques and technologies, along with the integration of Artificial Intelligence (AI) algorithms, have significantly contributed to the growth of this field. 

Particularly, Cone Beam Computed Tomography (CBCT) has become a powerful, cost-effective imaging technology for dental applications. It offers precise and accurate images while exposing patients to relatively low radiation doses \cite{cbct_tech}. In addition, its detailed three-dimensional reconstructions of the oral and maxillofacial regions can be used to diagnose several dental pathologies and conditions. Moreover, as an imaging technology, CBCT is not an exception when employing AI techniques for automatic diagnosis, and extensive research is being done in this line \cite{dent_pat_0}.

One common initial step in these algorithms is identifying and segregating teeth for their subsequent analysis. Typically, teeth are first located and identified. Then, a small bounding box containing the tooth and some context is used as input to another pipeline that infers other features \cite{diagnocat_2018, diagnocat_2022}.

We propose a tooth detection, identification, and reconstruction algorithm based on a three-stage pipeline. First, the three-dimensional CBCT image is divided into axial slices where the present teeth are located and identified with a one-stage object detector. Then, the two-dimensional bounding boxes are aligned, matched, and merged to generate three-dimensional tooth models. Ultimately, we would tackle artifacts due to the deficient teeth split. We treat the images as graphs to satisfactorily delineate teeth boundaries, with each voxel connected to its adjacent neighbors within the same sagittal slice. Through a sequence of heuristic-based penalties, we assign weights to each node. These weights help trace the interocclusal space division boundary. As a result, our method yields three-dimensional bounding boxes, each enclosing an individual tooth model along with a small surrounding context area around it.

The rest of the paper is organized as follows. Section \ref{sec:related_work} introduces other detection and segmentation methods, situating our proposal. Section \ref{sec:method_materials} explains the main blocks of our algorithm required to build the three-dimensional tooth models. Section \ref{sec:experiments} proposes several experiments to test our solution. Finally, Section \ref{sec:conclusion} concludes with future guidelines.

\section{Related work}
\label{sec:related_work}

Tooth identification and segregation are well-researched subjects. The latter generally comprises tooth segmentation or bounding box creation. Traditional methods often rely on features and structural extraction using thresholding, contour detection, or any other image transformations \cite{classical_2, classical_3}. These methods are convenient as they do not require an extensive dataset or expert knowledge for the labeling task. However, deep learning methods consistently outperform them \cite{on_tooth_segmentation}.

Dental fillings and other restorations may cause artifacts in CBCT images, leading to incorrect inferences. Deep learning-based methods usually rely on CNN architectures such as U-Net \cite{2d_unet}, V-Net \cite{3d_vnet}, and 3D U-Net \cite{3d_unet}, which are more suited to treat these imperfections. Moreover, these studies usually cover a broad range of tasks, from the location of the teeth to the identification of their labels in two-dimensional and three-dimensional slices.

Most three-dimensional approaches to interpreting CBCT images are based on multi-stage methodologies \cite{diagnocat_2018, seg_1, seg_2}. Furthermore, some studies segment the teeth from the rest of the CBCT image, extracting other structures such as bone tissue \cite{seg_3}. While deep learning-based segmentation approaches result in valuable features, they tend to need complex and relatively extensive datasets. Besides, operating deep learning models directly applied to three-dimensional images tends to be resource-intensive. These techniques may also need more flexibility when adapting to specific hardware configurations.

Conversely, alternative approaches perform inference on two-dimensional projections, such as axial slices or panoramic images. Nevertheless, these algorithms require a volume restoration stage in which the three-dimensional tooth is reconstructed \cite{tooth_detection_1, segmentation_in_axial}. Our method builds on this idea, including a custom algorithm to remedy artifacts. Overall, our approach offers a resource-efficient, fast, interpretable, and robust method. It significantly mitigates the effort and expert knowledge required for labeling, making it a practical solution in dental imaging applications.

\section{Method and materials}
\label{sec:method_materials}

The proposed method for solving the tooth detection, identification, and segregation problem consists of three steps:

\begin{enumerate}[1.]
\item \label{itm:first_ma} In the first stage, the images are divided into equispaced axial slices. A deep learning model detects and classifies the teeth in each axial slice into eight possible categories. Given the symmetry of dental structures and the resemblance between the mandible and maxilla in the axial dimension, these categories do not differentiate between the top and bottom or the right and left sides of the maxillofacial region. Following the FDI 2-digit World Dental Federation (ISO) notation \cite{iso_teeth}, each of our eight labels corresponds to the next teeth:

\begin{itemize}
    \item First incisor (1): 11, 21, 31, 41.
    \item Second incisor (2): 12, 22, 32, 42.
    \item Canine (3): 13, 23, 33, 43.
    \item First premolar (4):  14, 24, 34, 44.
    \item Second premolar (5): 15, 25, 35, 45.
    \item First molar (6): 16, 26, 36, 46.
    \item Second molar (7): 17, 27, 37, 47.
    \item Third molar (8): 18, 28, 38, 48.
\end{itemize}

\item \label{itm:second_ma} Then, individual three-dimensional bounding boxes containing each separate tooth are extracted. This step is formulated as an optimal assignment problem. Precisely, we use the Hungarian algorithm to match all the bounding boxes that belong to the same tooth. This approach is inspired by a tracking task \cite{tracking_1} along the axial slices.  

\item \label{itm:third_ma} Finally, some artifacts may emerge when upper and lower teeth lack separation, occluding the interdental space. To mitigate this, practitioners often employ a silicone bite block. This device is inserted into the mouth to stabilize and position the patient's jaws. However, even with a dental bite block, some instances may still lead to overlapped reconstructions. We have designed a graph-based approach to calculate the division of these distinctive cases.

\end{enumerate}

The pipeline inputs a three-dimensional CBCT image and produces a set of three-dimensional bounding boxes containing the identified teeth and some context around them. Eventually, these bounding boxes adapt tightly to circumvent teeth boundaries without sufficient interoclussal space.

\subsection{Dataset and notation}
\label{sec:dataset_and_notation}

Our dataset comprises 250 anonymized CBCT images displaying the maxillofacial region. It encompasses both natural teeth and dental prosthetics, including implants and bridges. For the purpose of our study, these prosthetics have been classified alongside natural teeth and have been assigned one of the eight possible labels.

As part of the training process for the detection model, 15 equispaced axial slices have been extracted from each CBCT image. The dataset has been randomly partitioned into training, testing, and validation subsets, adhering to an 80/10/10 distribution. 

\begin{table}[h]
\centering
\dcaption{\label{tab:tooth-detection-dataset}Tooth detection dataset}{Distribution of appearances per tooth class in the axial slices used to train and evaluate the detection model.}
\begin{tabular}{l|lll}
                    & Train & Validation & Test \\ \hline
First incisor (1)   & 4147  & 471        & 529  \\
Second incisor (2)  & 4212  & 480        & 547  \\
Canine (3)          & 4853  & 579        & 608  \\
First premolar (4)  & 3995  & 441        & 503  \\
Second premolar (5) & 3469  & 409        & 420  \\
First molar (6)     & 3207  & 375        & 447  \\
Second molar (7)    & 3376  & 414        & 445  \\
Third molar (8)     & 1428  & 156        & 179 
\end{tabular}
\end{table}

Table \ref{tab:tooth-detection-dataset} presents our dataset's distribution per tooth class. It is important to note that specific axial slices in the dataset may not contain any teeth. Furthermore, this distribution is influenced by the patient's number of teeth and size. These factors contribute to the slight discrepancies observed in the distributions across the three subsets.

\begin{figure}[h!t]
  \includegraphics[width=\columnwidth]{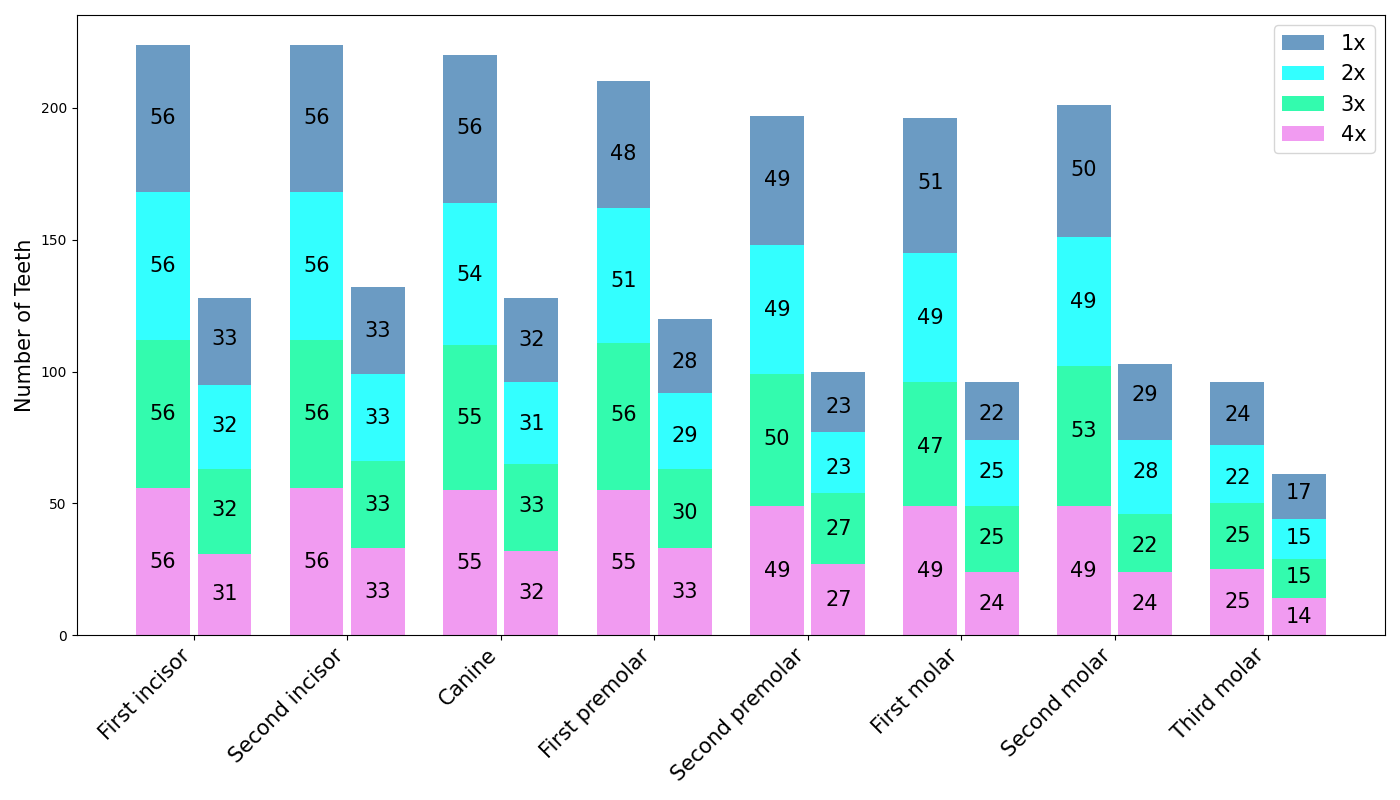}
  \dcaption{Tooth division and reconstruction dataset}{Distribution of the labels in FDI notation (in increasing sub-label order). There are 89 CBCT images, 56 with bite block (left) and 33 without bite block (right).}
  \label{fig:tooth-division-dataset}
\end{figure}

To evaluate the reconstruction and correction algorithms discussed in Sections \ref{sec:reconstruction_algorithm} and \ref{sec:double_teeth_correction}, we have selected a subset with 89 CBCT images. This dataset includes 56 images captured using a bite block to separate the mandible and maxilla regions, along with 33 images taken without this device. Following the FDI notation format, the distribution of teeth per class is detailed in Figure \ref{fig:tooth-division-dataset}. As previously observed in Table \ref{tab:tooth-detection-dataset}, incisors and canines are more frequent, while third molars are the least common. Nevertheless, the ordering is consistent, and the distribution patterns in both sets are similar. The bite block sub-division reflects a compatible prevalence of certain tooth types over others, balancing the dataset suitably.


\subsection{Tooth detection}
\label{sec:detection}

The first step of our algorithm involves detecting and identifying teeth in axial slices. For this purpose, we have trained a single-stage deep learning model based on a fine-tuned version of YOLOv7 \cite{yolo_v7}. This model localizes the teeth and identifies one of the eight possible labels described in Section \ref{sec:dataset_and_notation}. The detection algorithm consists of two steps:

\begin{enumerate}[1.]
    \item In the analysis of CBCT images of the mouth, axial slices are first extracted at regular, equispaced intervals of 1.4 mm. This interval is chosen based on the minimum expected size of an adult tooth, ensuring at least three axial slices per tooth. Furthermore, the region from which the teeth are extracted is identified as the image section that displays the highest mean axial value when projected to the vertical axis. More precisely, during our experiments, we have found that it is reasonable to define this region as the window that displays the top 90\% range of mean axial slice values upon projection to the vertical axis. It is important to note that this interval of 1.4 mm between axial slices is applied during inference. The training dataset does not adhere to the 1.4 mm interval between consecutive axial slices; instead, 15 axial slices per image are extracted.
    \item The model evaluates each axial slice, resulting in a collection of two-dimensional detections. Each detection is associated with its axial slice (enumeration identifier) and a label.
\end{enumerate}

\begin{figure}[h!t]
  \includegraphics[width=0.3\textwidth]{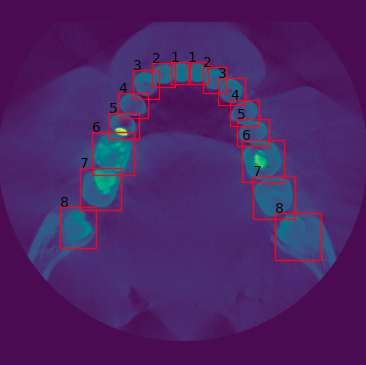}
  \dcaption{Tooth detection in an axial slice}{No distinction between top and bottom, or right and left sides.}
  \label{fig:results_yolo}
\end{figure}

One instance of the detections predicted by our model is depicted in Figure \ref{fig:results_yolo}. This image shows an axial slice of the mandible, where the teeth on the right side of the mouth are assigned the same label as those mirrored on the left.

\subsection{Tooth reconstruction}
\label{sec:reconstruction_algorithm}

\begin{figure*}[h!t]
  \includegraphics[width=1\textwidth]{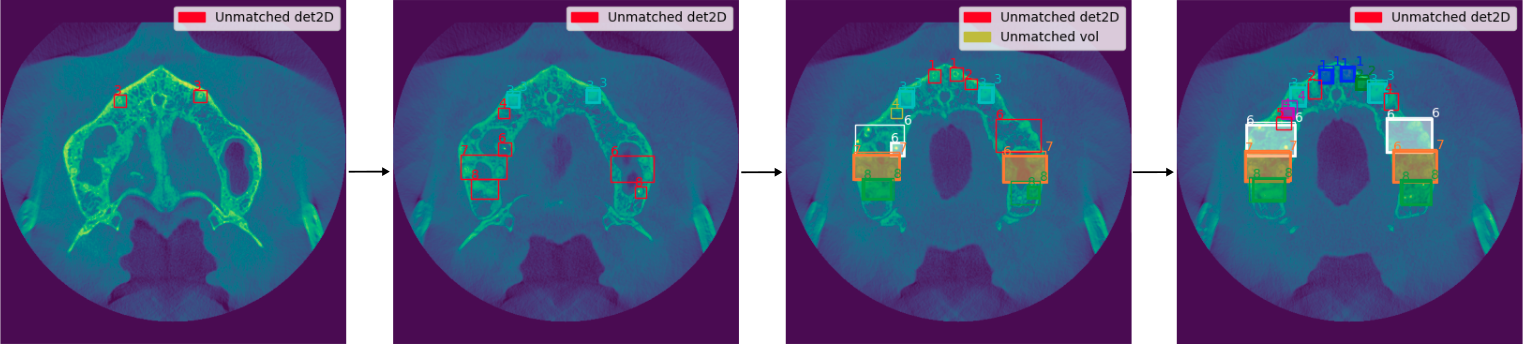}
  \dcaption{Matches in consecutive axial slices}{In red, the tooth volumes under construction with a single match. In yellow, the tooth volumes under construction with no matches in that axial slice. In other colors, the "active" tooth volumes (filled) matched with a new detection (not filled) in the axial slice under study.}
  \label{fig:matching_algorithm}
\end{figure*}

Building upon the previous detections in axial slices, we construct the candidate volumes that contain each tooth. Essentially, the volumes consist of a collection of two-dimensional bounding boxes. This step involves iteratively aligning and joining the slices corresponding to the same tooth. This process begins by analyzing the top axial slice that contains detections and proceeds downward until all axial slices of the CBCT image have been analyzed. The following metrics are employed to evaluate the quality of a new match:

\begin{itemize}

    \item We contrast between a bounding box in the axial slice under study and a previously matched bounding box in another axial slice. The latter corresponds to the last match within the tooth volume under construction. The Euclidean distance in the vertical plane is used to measure proximity. It facilitates the identification of detections across non-adjacent axial slices. Furthermore, a limit can be set on the maximum number of consecutive axial slices where a tooth is present but not detected (i.e., missed detections). Beyond this limit, any new match is deemed invalid, resulting in the tooth being fully reconstructed.

    \item The Intersection over Union (IoU) measures the overlap between a bounding box on the current axial slice, and the projection of the last match of the volume under construction onto the same axial slice.
    
    \item A minor penalty is imposed when a bounding box and a tooth volume with differing labels are matched. The tooth volumes under construction may potentially belong to multiple classes simultaneously. This situation arises because a label is assigned to a tooth volume based on its bounding box collection. Therefore, the penalty for label mismatch varies in proportion to the uncertainty of the label of the tooth volume being constructed. The higher the uncertainty of the match, the greater the weight of the penalty. This approach allows matching bounding boxes that belong to the same tooth, even when the detection algorithm mislabels them. Although the detection model is proficient at locating teeth, it occasionally mislabels some detections.
    
\end{itemize}

During construction, tooth volumes are labeled as "active" or "closed". "Active" indicates tooth volumes that are still under construction and can accept further matches, whereas "closed" refers to tooth volumes that have been fully reconstructed. The matching algorithm operates as follows:

\begin{enumerate}[1.]
    
    \item Start with the top axial slice containing detections. Create a tooth volume for each detection and mark them as "active".
    \item Continue scrutinizing the successive axial slices until exhausted. For each new axial slice:

    \begin{enumerate}[2.1.]
        \item \label{step:2.1-reconstruction} Calculate the optimal match between the new detections and the "active" tooth volumes under construction. This process involves weighting the metrics described above, and utilizing the Hungarian algorithm to infer the optimal match. Additionally, an upper threshold beyond which a match between a two-dimensional detection and a tooth volume is not permitted can be imposed. We have achieved favorable results by setting this threshold such that, if there is no overlap in the axial dimension, the match is deemed invalid.
        \item Create a new "active" tooth volume for those detections in the axial slice under study that have not been matched with any existing "active" tooth volumes.
        \item Tooth volumes that have not had matches in a specific number of iterations are marked as "complete" and will not accept further matches. The label assigned to a tooth is determined by the most frequent among the corresponding two-dimensional detections. In our experiments, setting the maximum number of skipped axial slices without a detection to two has yielded good results.
    \end{enumerate}
    
    \item Teeth not detected in at least three axial slices are considered invalid. This condition makes the algorithm more robust to further false detection errors, primarily due to odd bone structures, often mistaken for teeth roots. This number assesses both statistical relevance and empirical experience, as tooth volumes smaller than 2.8 mm are irrelevant as far as we are concerned.    
\end{enumerate}

Figure \ref{fig:matching_algorithm} illustrates an example of matches across consecutive axial slices. This figure demonstrates the generation of new tooth volumes using unmatched two-dimensional bounding boxes (depicted in red). As well, other two-dimensional bounding boxes (in various colors) are matched with their corresponding tooth volumes. Notably, even though the left tooth labeled as $4$ is not detected in the third axial slice, it is correctly matched in the fourth axial slice with the corresponding tooth volume of the second axial slice. Additionally, the right tooth, incorrectly labeled as $6$ in the second axial slice, is accurately matched with its detections labeled as $7$ in both the third and fourth axial slices.

\begin{figure}[h!t]
   \centering
   \subfloat{
   	\label{fig:3d_reconstruction-real}
   	\includegraphics[width=0.2\textwidth]{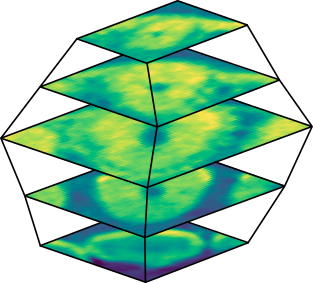}} 
  \caption{Matching two-dimensional bounding boxes to model a three-dimensional volume containing a tooth.}
  \label{fig:3d_reconstruction}
\end{figure}

\begin{figure}[h!t]
\centering

\subfloat[Tooth 32]{
	\label{fig:tooth_32}
	\includegraphics[width=0.24\textwidth]{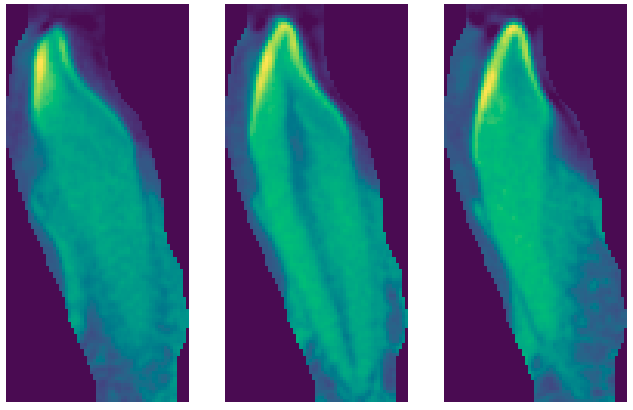}} 
 
\subfloat[Tooth 23]{
	\label{fig:tooth_23}
	\includegraphics[width=0.24\textwidth]{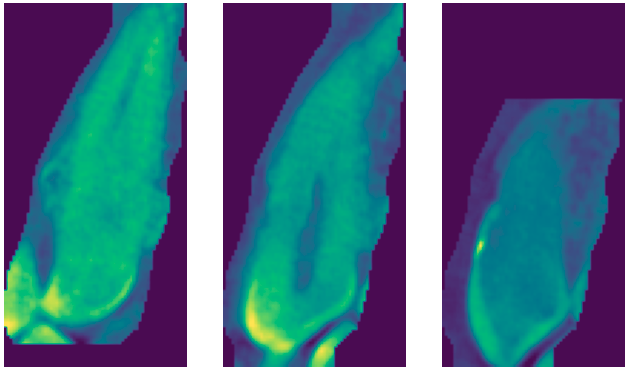}} 
 
\subfloat[Tooth 37]{
	\label{fig:tooth_37}
	\includegraphics[width=0.24\textwidth]{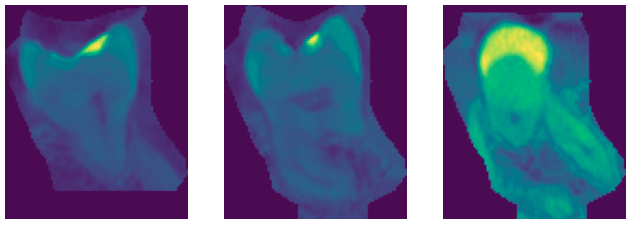}} 
 
\dcaption{Teeth reconstruction}{Sagittal (a and b) and coronal (c) views of the teeth.}
\label{fig:tooth_reconstruction_example}
\end{figure}

The tooth volumes created with matched detections are sparsely distributed. The sparse distribution arises because the detection model is only applied to a subset of the axial slices to meet computational requirements. Consequently, the gap between matched detections is filled using linear interpolation. Figures \ref{fig:3d_reconstruction} and \ref{fig:tooth_reconstruction_example} illustrate this process. The former depicts the sparsely distributed bounding boxes and the linear interpolation between their edges. The latter presents three examples of the results from this interpolation. 


Moreover, as explained in step \hyperref[step:2.1-reconstruction]{2.1} of the matching algorithm, the metrics must be weighted as follows:

\begin{equation}
\label{eq:tooth_reconstruction}
    \begin{split}
        q(t, b^l) & = w^{(1)}_{q} \cdot d(t,b^l) \cdot h(t, b^l) \\
               & + w^{(2)}_{q} \cdot (1 - IoU(t, b^l)) \\
               & + w^{(3)}_{q} \cdot \left(1 - f_{l}(t)\right) \\
    \end{split}
\end{equation}

\noindent where:

\begin{itemize}
    \item $t$: Tooth volume under construction. It consists of an ordered collection of bounding boxes ($b^l$).
    \item $b^l$: Candidate bounding box with label $l$.
    \item $l$: The label assigned to a bounding box. This parameter can take one out of eight possible values:  $l \in \{1, 2, 3, 4, 5, 6, 7, 8\}$.
    \item $q(t,b^l)$: Quality of a match between a tooth volume under construction $t$ and a new bounding box $b^l$. A lower value of $q(t,b^l)$ indicates a better match.
    \item $d(t,b^l)$: One-dimensional Euclidean distance along the vertical axis between the candidate bounding box $b^l$ and the closest component of $t$.
    \item $h(t, b^l)$: Function that sets the corresponding component of the equation to zero if the distance is less or equal to the configurable hyperparameter $\gamma$, which sets a proximity lower bound. It is defined as:
    
    \begin{equation}
    \label{eq:step_distance}
        \begin{split} 
        & h(t, b^l) = 
            \begin{cases} 
            0 & \text{if } d(t,b^l) \leq \gamma \\
            1 & \text{if } d(t,b^l) > \gamma \\
            \end{cases}
        \end{split}
    \end{equation}
        
    \item $IoU(t,b^l)$: Intersection over union metric applied to the axial projections of the bounding boxes.    
    \item $f_{l}(t)$: Percentage of bounding boxes already matched with the tooth volume $t$ that were identified with the label $l$.


    \item $w^{(1)}_{q}$, $w^{(2)}_{q}$ and $w^{(3)}_{q}$: Weights associated with each of the metrics. 
    
\end{itemize}

In the aforementioned equation, a new match will be discarded if $q(t,b^l) > \beta$. The boundary condition $\beta$ is a configurable parameter that serves as an upper threshold. It must be set to a fixed number, and the remaining weights must be calibrated accordingly.

Through our experiments, we have found that the following configuration of parameters is reasonable:

\begin{itemize}
    \item $w^{(2)}_{q} \gtrsim \beta$: This configuration ensures that if $b^l$ does not overlap in the axial plane with $t$, the match will be disregarded. In practice, $w^{(2)}_{q} \approx \beta$ yields good results.
    \item $\gamma$: We have defined $\gamma$ as the Euclidean distance of the step size between adjacent axial slices. This minimum distance ensures that the first term of Equation \ref{eq:tooth_reconstruction} does not penalize a match between consecutive axial slices.
    \item $w^{(1)}_{q} = \frac{\beta}{4 \cdot \gamma}$: This parameter establishes the maximum number of consecutive axial slices allowed without detecting $t$. We have achieved the most favorable results by relaxing this condition and computationally setting the limit of maximum axial slices allowed without a detection of $t$ to two. According to our setup, if $t$ is detected in the previously analyzed axial slice, no penalty associated with the first term of Equation \ref{eq:tooth_reconstruction} is applied. However, skipping one axial slice results in a penalty of $1/2 \cdot \beta$, and skipping more than one axial slice disqualifies the match.
    \item $w^{(3)}_{q} \approx 0.2 \cdot \beta$: Regarding the label mismatch, we have found that it is not a critical term, and its weight can be configured as a fraction of the maximum allowed penalty $\beta$.
\end{itemize}

\subsection{Artifacts and correction}
\label{sec:double_teeth_correction}

\begin{figure}[h!t]
\centering

\subfloat[]{
	\label{fig:double_tooth}
	\includegraphics[width=0.08\textwidth]{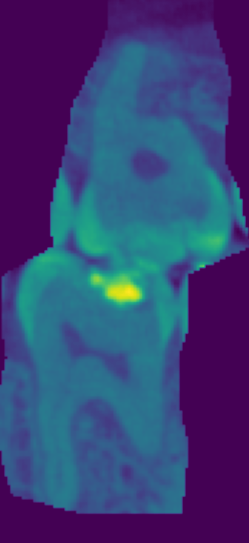}}
\qquad
\subfloat[]{
	\label{fig:t_sep_sup}
	\includegraphics[width=0.08\textwidth]{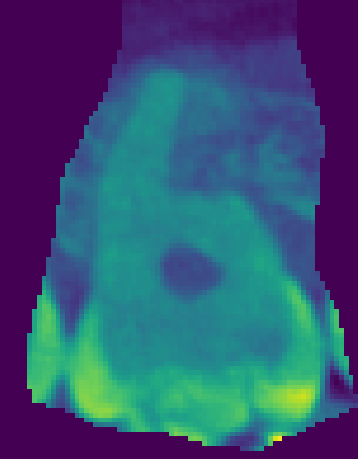}} 
\qquad
\subfloat[]{
	\label{fig:t_sep_inf}
	\includegraphics[width=0.08\textwidth]{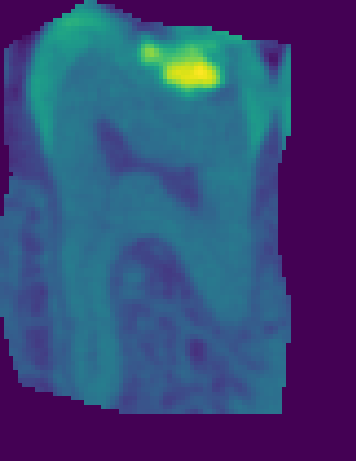}} 
\dcaption{Occlusal counterparts (a) and divided teeth (b, c)}{The division performed by our algorithm is adapted to the actual geometry of the space between the two teeth.}
\label{fig:examples_teeth}
\end{figure}

When insufficient interocclusal space exists, sequential volume construction may fail. Even with a bite block, there is a potential risk of joining the occlusal counterparts (Figure \ref{fig:double_tooth}). A heuristic-based, graph-oriented strategy ensures accurate separation, even in cases where the division of the mandible and maxilla regions is challenging, primarily due to the limited interocclusal space. Our algorithm effectively separates these fused counterparts into distinct entities (Figures \ref{fig:t_sep_sup} and \ref{fig:t_sep_inf}).

Firstly, we apply "if/else" rules to detect these artifacts automatically. Joined teeth will be significantly larger than any other tooth from the same scan. We compute the size of each tooth and flag volumes that deviate significantly from the expected size distribution for review.

Subsequently, those volumes containing two teeth are processed using our graph-based division algorithm. In CBCT images, voxel values are directly proportional to their density \cite{hu_units}. Ideally, the projection of the mean value of each axial slice onto the vertical axis will delineate two prominent peaks representing each tooth, with an intervening valley corresponding to the mouth aperture. This statement remains true, provided there is sufficient interocclusal space. In cases where this space is minimal, we computationally infer its expected position, typically at the midpoint of the volume. This estimate supports the search process as an initial condition. Manifold penalties related to the density of the pixels and boundary conditions confine the results in the desired range of values. Finally, all the identified divisions are combined into a lattice that adequately separates the volumes.

\section{Experiments}
\label{sec:experiments}

In this section, we present the experiments conducted to assess each of the three components of our algorithm. First, we evaluate the detection model described in Section \ref{sec:detection}. Subsequently, we examine the accuracy of the reconstruction algorithm explained in Section \ref{sec:reconstruction_algorithm}. Finally, we test the division algorithm discussed in Section \ref{sec:double_teeth_correction}.

The detection model utilized in our study is based on the YOLO architecture. We carried out a series of fine-tuning procedures on various YOLO versions, achieving the most satisfactory results with YOLOv7. Our final model was trained using 3,375 axial images from the training and evaluation sets and tested on a separate set of 375 images. For training, we used the Adam optimizer for 200 epochs. All axial images were resized to $416 \times 416$. We set the object confidence threshold and the IoU threshold for non-maximum suppression at 0.4. The resulting model achieved a Mean Average Precision at an IoU threshold of 0.5 (mAP@0.5) of 0.889, and a Mean Average Precision at IoU thresholds ranging from 0.5 to 0.95 (mAP@0.5:0.95) of 0.606.  

It should be emphasized that the reconstruction algorithm is tolerant to some errors in both the label assigned to the detections and the spurious detections. This tolerance stems from the algorithm's reliance on heuristics that evaluate the label and the detection's position.

Table \ref{tab:map-yolo} displays the mAP@0.5 and mAP@0.5:0.95 results for each tooth class. The molars, especially the third molars, present the most influential challenge for labeling. This is due to their similarity in the axial dimension, the limited number of examples of this class, and the more outstanding variability in their positions compared to the frontal teeth. Nevertheless, given that the reconstruction algorithm tolerates some errors at this stage, the results are considered satisfactory.

\begin{table}[h!t]
\centering
\caption{\label{tab:map-yolo}mAP of the detection model per tooth class.}
\begin{tabular}{l|lll}
    & \# Labels & mAP@.5 & mAP@.5:.95 \\
\hline

First incisor (1) & 529 & 0.940 & 0.602 \\
Second incisor (2) & 547 & 0.921 & 0.591 \\
Canine (3) & 608 & 0.961 & 0.663 \\
First premolar (4) & 503 & 0.905 & 0.602 \\
Second premolar (5) & 420 & 0.877 & 0.601 \\
First molar (6) & 447 & 0.819 & 0.598 \\
Second molar (7) & 445 & 0.863 & 0.629 \\
Third molar (8) & 179 & 0.824 & 0.558 \\

\end{tabular}
\label{table:map_results}
\end{table}

To assess the performance of the reconstruction and division algorithms, we compiled a dataset consisting of 89 CBCT images. This dataset was divided into two subsets: the first subset comprises 56 CBCT images in which a bite block is used, and there is some interocclusal space. In contrast, the second subset includes 33 CBCT images where the bite block was not utilized, and there is no interocclusal space. The first subset encompasses a total of 1568 teeth, and the second includes 868 teeth.

The rationale behind dividing the dataset into two subsets was motivated by the significant influence of interocclusal space on the results. First, the reconstruction algorithm tends to create fewer volumes containing two teeth when there is a slight separation between the mandible and maxilla regions. Moreover, more favorable outcomes are observed in the subset with interocclusal space in cases where the division algorithm separates a volume with two teeth. These better results are achieved because the inferred division is less constrained and can adapt closely to the teeth's geometry.

Our algorithm successfully detected and reconstructed all but 26 teeth, resulting in a detection rate of 98.93\% (this value does not necessarily reflect accurate reconstruction). Most detection errors were attributed to small dental bridges and implants, frequently facing misclassification issues caused by reflections and suboptimal image quality. Additionally, the detection model mistakenly identified some bone structures as tooth roots. Nevertheless, our reconstruction algorithm employs a minimum size criterion of $2.8mm$ for a volume to be accepted as a tooth. Consequently, there was only one wrong model (false positive) in the dataset of 89 CBCT images, where a small bone structure in the nostril area was incorrectly classified and modeled as a tooth root.

We have designed two tests to evaluate the reconstruction and division algorithms independently. First, we classify the results of the reconstruction algorithm (prior to division) into the next categories:

\begin{itemize}
    \item Single tooth: The volume contains a single tooth.
    \item Double tooth: The volume contains two teeth (before the division algorithm is applied).
    \item Not detected: The tooth is present, but our algorithm does not detect it.
\end{itemize}

This test does not aim to evaluate perfect teeth modeling, where all tooth voxels are accurately captured. Instead, it serves as a foundational metric to estimate the frequency of cases requiring the application of the division algorithm. The results are presented in Table \ref{tab:pre-div-metrics}.

\begin{table}[h!t]
\centering
\caption{\label{tab:pre-div-metrics}Results before applying the division algorithm.}
\begin{tabular}{l|ll}
             & With & Without  \\ 
             & occlusal clearance & occlusal clearance  \\ 
\hline
Single tooth & 1194 (76.15\%)     & 464 (53.46\%)      \\
Double tooth & 178 (22.70\%)      & 198 (45.62\%)      \\
Not detected & 18 (1.15\%)        & 8 (0.92\%)
\end{tabular}
\end{table}

In addition, we have evaluated the combination of the reconstruction and division algorithms. Specifically, we applied the division algorithm to the reconstructed volumes containing two teeth. Subsequently, we categorized the outcomes into four distinct groups:

\begin{itemize}
    \item Good reconstruction: The volume contains all the voxels belonging to a single tooth, along with some additional context (a few additional voxels around the tooth).
    \item Bad reconstruction: The volume contains a single tooth. However, some voxels of the tooth were left outside of the volume.
    \item Double tooth: This volume contains two teeth but was identified as if it contained a single tooth. Consequently, the volume still includes both teeth.
    \item Not detected: The tooth is present, but our algorithm does not detect it.
\end{itemize}

The results of this second test are detailed in Table \ref{tab:post-div-metrics}. There is a notable disparity when our algorithm is applied to CBCT images with some interocclusal space compared to those with complete occlusion. The presence of a gap between the teeth reduces the need to use the division algorithm, as shown in Table \ref{tab:pre-div-metrics}. Additionally, edge computation is significantly simplified in cases where there is some space between the mandible and maxilla regions. In contrast, the division must conform more closely to the teeth's geometry in cases without this small space.

\begin{table}[h!t]
\centering
\caption{\label{tab:post-div-metrics}Reconstruction performance.}
\begin{tabular}{l|ll}
             & With & Without  \\ 
             & occlusal clearance & occlusal clearance  \\ 
\hline
Good reconstruction & 1509 (96.24\%)    & 743 (85.60\%)    \\
Bad reconstruction  & 33 (2.10\%)       & 87 (10.02\%)     \\
Double tooth  & 4 (0.51\%)              & 15 (3.46\%)      \\
Not detected & 18 (1.15\%)              & 8 (0.92\%)
\end{tabular}
\end{table}

\section{Discussion and conclusion}
\label{sec:conclusion}

We have successfully designed and developed an algorithm for detecting and identifying teeth in three-dimensional CBCT images. The initial phase of our algorithm focuses on detecting teeth in two-dimensional axial slices. This approach simplifies the detection and identification process by requiring less complex training data. Furthermore, it makes the annotation process more straightforward and accessible.

We have created a fast and reliable automatic process for generating three-dimensional teeth models by integrating this streamlined detection method with a proven reconstruction and division algorithm. These models can be used directly by experts or serve as a foundation for further segmentation techniques. For instance, the segmented three-dimensional teeth can be utilized in various tasks, including pathology detection, demonstrating the algorithm's versatility and effectiveness in dental imaging.

The primary challenge lies in detecting and dividing reconstructed tooth volumes containing two teeth, especially in scenarios with minimal interocclusal space. One potential solution to improve the inference in such cases is to employ another deep learning model to detect both the volumes that contain two teeth and the approximate height of the division. In addition, density-based erosion methods could further refine the teeth volumes for precise pixel-wise segmentation.

Multiple domain experts have confirmed the effectiveness and reliability of our solution. Moreover, our detection and identification algorithm has been integrated into the tooth analysis tool Dentomo \cite{dentomo}, improving expert efficiency by simplifying and reducing diagnosis time.


\bibliographystyle{ACM-Reference-Format}
\bibliography{bib_samples}

\end{document}